\documentclass[prb,showpacs,amsmath,amssymb,superscriptaddress,
floatfix,twocolumn]{revtex4}

\usepackage[english]{babel}
\usepackage{graphicx}
\usepackage{times}
\usepackage{units}

\begin{document}

\title{Properties of anisotropic magnetic impurities on surfaces}

\author{R. \v{Z}itko}
\affiliation{Institute for Theoretical Physics, University of G\"ottingen,
Friedrich-Hund-Platz 1, D-37077 G\"ottingen, Germany}
\affiliation{J. Stefan Institute, Jamova 39, SI-1000 Ljubljana, Slovenia}

\author{R. Peters}
\affiliation{Institute for Theoretical Physics, University of G\"ottingen,
Friedrich-Hund-Platz 1, D-37077 G\"ottingen, Germany}

\author{Th. Pruschke}
\affiliation{Institute for Theoretical Physics, University of G\"ottingen,
Friedrich-Hund-Platz 1, D-37077 G\"ottingen, Germany}

\date{\today}

\begin{abstract}
Using numerical renormalization group techniques, we study static and
dynamic properties of a family of single-channel Kondo impurity models with
axial magnetic anisotropy $DS_z^2$ terms; such models are appropriate to
describe magnetic impurity atoms adsorbed on non-magnetic surfaces, which
may exhibit surface Kondo effect. We show that for positive anisotropy $D$
and for any spin $S$, the systems behave at low temperatures as regular
Fermi liquids with fully compensated impurity spin. The approach to the
stable fixed point depends on the value of the spin $S$ and on the ratio
$D/T_K^{(0)}$, where $T_K^{(0)}$ is the Kondo temperature in the absence of
the anisotropy. For $S=1$, the screening occurs in two stages if
$D<T_K^{(0)}$; the second Kondo temperature is exponentially reduced in this
case. More generally, there is an effective spin-$1/2$ Kondo effect for any
integer $S$ if $D<T_K^{(0)}$ and for any half-integer $S$ if $D>T_K^{(0)}$.
For negative anisotropy $D$, the system is a non-Fermi liquid with residual
anisotropic exchange interaction. However, the presence of transverse
magnetic anisotropy $E(S_x^2-S_y^2)$ restores Fermi-liquid behavior in real
systems.
\end{abstract}


\maketitle

\newcommand{\vc}[1]{{\boldsymbol{#1}}}
\newcommand{\ket}[1]{|#1\rangle}
\newcommand{\bra}[1]{\langle #1|}
\newcommand{\braket}[1]{\langle #1 \rangle}
\renewcommand{\Im}{\mathrm{Im}}
\renewcommand{\Re}{\mathrm{Re}}
\newcommand{\dr}{\mathrm{d}}
\newcommand{\correl}[1]{\langle\langle #1 \rangle\rangle_\omega}
\newcommand{\TKO}{T_K^{(0)}}
\newcommand{\TKt}{T_K^{(2)}}

\newcommand{\figw}{7cm}
\newcommand{\cchi}{\mathcal{S}}

\section{Introduction}

As the sizes of the active components in magnetic devices shrink, the ratio
of the surface area over bulk volume steadily increases and the importance
of surface and interface effects grows. In the past, the focus has mostly
been on the magnetic properties of clean surfaces, thin films, and
interfaces in multilayers \cite{wiesendanger1992, farle1998, heinze2000,
vaz2008}, but now magnetism of nanostructures is emerging as a new subfield
with the aim of exploring and modifying the exchange interaction at the
nanometer level. Given that the symmetry in the surface area is always
reduced as compared to the bulk, more attention needs to be given to
possible additional surface-induced magnetic anisotropy effects
\cite{heinze2000, gambardella2002, rusponi2003, kuch2003, heinrich2004,
bode2007, hirjibehedin2007, ferriani2008, vaz2008}. Recently, magnetic
anisotropy on surfaces studied at the atomic level has become a new focus of
interest with potential applications in ultra-dense magnetic data storage.
Using scanning tunneling spectroscopy, the magnetic anisotropy can be
measured on the single-atom level by analyzing the splittings of magnetic
excitation peaks as a function of the applied magnetic field
\cite{heinrich2004, hirjibehedin2007}. Magnetic anisotropy constants were
found to be very large, often in the $\unit{meV}$/atom range
\cite{gambardella2002, gambardella2003, hirjibehedin2007}.

Single magnetic impurity atoms adsorbed on surfaces or buried in the
near-surface regions can be described using quantum impurity models
\cite{hewson, bulla2008} with anisotropy terms \cite{ujsaghy1996,
hirjibehedin2007}. The dominant anisotropy terms are of the form $D (\vc{S}
\cdot \vc{n} )^2$, where $\vc{n}$ is a vector along a privileged axis that
is commonly chosen to be the $z$-axis \cite{ujsaghy1996, ujsaghy1999}. The
most important contributions seem to arise from the local spin-orbit
coupling, as the reduced symmetry leads to unequal hybridization of the
different $d$-orbitals of the impurity atom \cite{szunyogh2006, szilva2008}.
For an impurity buried in the near-surface region, $D$ depends on the
distance from the surface and oscillates in sign \cite{szilva2008}. 

Numerical calculations and experiments show that the magnetic anisotropy
strongly depends on the microscopic details. Depending on the impurity
species and on the substrate, different directions of the privileged axis
may be found. Density functional theory calculation predict, for example,
that perpendicular anisotropy is common for single adatoms on metallic
surfaces \cite{etz2008}.

Spin-excitation spectroscopy using a scanning tunneling microscope (STM) has
revealed a rich variety of different classes of behavior. For example, for
Fe (spin-$2$) adatom on a thin CuN decoupling layer on a Cu surface, the
anisotropy is negative with the $z$-axis in the surface plane, along the
direction of nitrogen atoms, while for Mn (spin-$5/2$) on the same surface,
the anisotropy is weaker but with an out-of-plane $z$-axis, however $D$ is
still negative \cite{hirjibehedin2007}. No anomalies at the Fermi level
(surface Kondo effect \cite{li1998, madhavan1998, ujsaghy2000, schiller2000,
madhavan2001}) are observed for $D<0$ \cite{hirjibehedin2007}. For Co
(spin-$3/2$) on CuN layer, the anisotropy $D$ is positive and signatures of
the Kondo effect are observed in tunneling spectra \cite{heinrichprivate}.
These results strongly motivate a study of the entire family of the Kondo
impurity models for various values of spin and for different signs and
magnitudes of the magnetic anisotropy. Due to the low symmetry at the
surface, it appears sufficient as a first approximation to consider coupling
to a single conduction channel (the one with the largest exchange coupling
constant); it is unlikely that coupling constants to two orthogonal channels
would be equal under these conditions.

Similar impurity models with large anisotropies also appear in the context
of single molecular magnets, e.g. Mn$_{12}$ or Fe$_8$. It was possible to
attach metallic contacts to these molecules allowing for electron transport
measurements \cite{heersche2006,jo2006}. The molecules exhibit very large
spin values ($S=10$ for Mn$_{12}$) and a negative-$D$ axial anisotropy term
$D S_z^2$, so that ground states with two polarized large-spin states are
favored in the absence of quantum spin tunneling due to the transverse
anisotropy $E(S_x^2-S_y^2)$ terms \cite{romeike2006, romeike2006b,
roosen2008, bogani2008}.

Since the ultimate miniaturization limit appears to be storing information
in the spin states of single magnetic atoms, it is important to study
decoherence by possible magnetic screening effects through nearby conduction
electrons (Kondo effect), since the goal is to have long-lived large-spin
magnetic impurity states. The appropriate model is thus the impurity Kondo
model with additional anisotropy terms that we introduce in
Sec.~\ref{secmodel}. In Sec.~\ref{secscaling} we relate these models by
means of scaling equations to the Kondo models with XXZ anisotropic exchange
coupling to the conduction band. Numerical renormalization group results are
presented in Sec.~\ref{secresults} where we compare calculations for $S=1$,
$S=3/2$ and $S=2$. $S=3/2$ is a typical representative for all
half-integer-spin models, $S=2$ a typical representative for all
integer-spin models, while $S=1$ is a special case with some additional
features when compared to higher integer spins. We present results both for
static properties (magnetic susceptibility, impurity entropy) and for
dynamic properties (spectral functions, dynamic magnetic susceptibility).
This section also contains a discussion of the various fixed points,
focusing on the non-Fermi-liquid behavior in the $D<0$ models.

\section{Model}
\label{secmodel}

A general approach to study the anisotropy effects of adsorbed impurity
atoms would start with a multi-orbital Anderson model for the $d$-orbitals
with a suitable hybridization function for the coupling with the substrate
and with electron-electron repulsion and Hund coupling parameters extracted,
for example, from density functional theory calculations. The next step
would then consist of obtaining an effective Kondo-like model using a
Schrieffer-Wolff transformation. We aim, however, to explore only the
effects due to local magnetic anisotropy effects. We thus study a very
simplified Kondo impurity model defined by the following Hamiltonian:
\begin{equation}
\begin{split}
H &= H_\mathrm{band} + H_\mathrm{K} + H_\mathrm{aniso}, \\
H_\mathrm{band} &= \sum_{\vc{k}\sigma} \epsilon_{\vc{k}} 
c^\dag_{\vc{k}\sigma} c_{\vc{k}\sigma}, \\
H_\mathrm{K} &= J \vc{s} \cdot \vc{S}. \\
\end{split}
\end{equation}
Here operators $c_{\vc{k}\sigma}$ describe the conduction band electrons
with momentum $\vc{k}$ and spin $\sigma \in \{\uparrow, \downarrow\}$. In
numerical calculations, we will for simplicity assume a flat band
$\epsilon_{\vc{k}}=W k$, where the dimensionless momentum (energy) $k$ ranges
from $-1$ to $1$ and $W$ is the half-bandwidth; the density of states is
thus constant, $\rho=1/(2W)$. Furthermore, the effective exchange constant
$J$ is defined as 
\begin{equation}
J = \frac{1}{N^2} \sum_{\vc{k}\vc{k}'} J_{\vc{k},\vc{k}'},
\end{equation}
where $J_{\vc{k},\vc{k}'}$ describe exchange scattering of the conduction
band electrons on the impurity, and the spin-density of the conduction band
electrons $\vc{s}$ is defined as
\begin{equation}
\vc{s} = \frac{1}{N} 
\sum_{\vc{k}\vc{k}'\alpha\alpha'} 
\frac{J_{\vc{k},\vc{k}'}}{J}
c^\dag_{\vc{k}\alpha} 
\left( \frac{1}{2} \boldsymbol{\sigma}_{\alpha\alpha'} \right) 
c_{\vc{k}'\alpha'}.
\end{equation}
If $J_{\vc{k},\vc{k}'} \equiv J$ is constant, then $\vc{s}$ can be
interpreted as the spin-density of conduction band electrons at the position
of the impurity. For an impurity adsorbed on a surface, this clearly cannot
be a good approximation as the exchange interaction $J_{\vc{k},\vc{k}'}$ is
inevitably anisotropic in the reciprocal space; nevertheless, since we focus
mostly on the role of the local magnetic anisotropy terms, we will not
pursue this issue in this work. We furthermore assumed that the exchange
interaction is isotropic in the spin space.

\begin{figure}
\centering
\includegraphics[clip,width=8cm]{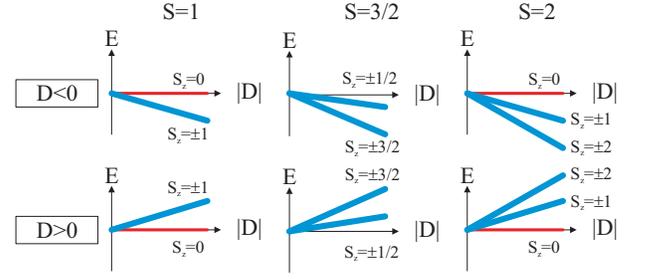}
\caption{(Color online) Energy level diagrams for the anisotropic
magnetic impurities with spin $1$, $3/2$ and $2$
as a function of the absolute value of the anisotropy,
$|D|$. Wide (blue) lines represent pairs of states,
narrow (red) lines represent non-degenerate states.}
\label{figa}
\end{figure}

In the anisotropy term of the Hamiltonian,
\begin{equation}
H_\mathrm{aniso} = D S_z^2 + E \left( S_x^2-S_y^2 \right),
\end{equation}
the first term is the {\sl axial magnetic anisotropy}, while the second is
named the {\sl transverse magnetic anisotropy}. The axis $z$ is by
convention chosen such that $|D|$ is maximized, while axes $xy$ are oriented
so that $E>0$. If the axial anisotropy term is negative, $D<0$, we call the
$z$ axis an {\sl easy axis}; if, however, $D>0$ the anisotropy is said to be
{\sl hard-axis} (or planar). Axial anisotropy term $D$ leads to the
splitting of magnetic levels, Fig.~\ref{figa}. In the case of easy-axis
anisotropy, $D<0$, the moment tends to be maximal in size with two
degenerate states pointing in the opposite directions, i.e. $S_z=\pm S$.
Since the conduction band electrons can only change the impurity spin by one
unit during a spin-flip scattering event, the exchange scattering rate is
expected to be strongly reduced for large enough $|D|$. In the case of
hard-axis anisotropy, $D>0$, the single level $S_z=0$ is favored for integer
$S$ and the doublet $S_z=\pm 1/2$ for half-integer $S$; in both cases, the
impurity moment will be compensated at low temperatures, since the doublet
can be screened via the conventional spin-$1/2$ Kondo screening. It should
be noted that $E(S_x^2-S_y^2) = E/2 \left[ (S^+)^2 + (S^-)^2 \right]$ term,
i.e. the transverse anisotropy, mixes levels with different values of $S_z$.
This term, even when $E$ is small, plays an essential role in the easy-axis
case \cite{romeike2006}; in the hard-axis case, however, it only leads to a
small correction.

The isotropic Kondo model (i.e. the $D=E=0$ limit) has been intensively
studied by a variety of techniques, such as numerical renormalization group
\cite{cragg1979b, cragg1980, mehta2005, koller2005}, Bethe Ansatz
\cite{furuya1982, andrei1983, andrei1984, andrei1994, sacramento1989,
sacramento1991td}, and other methods \cite{nozieres1980, coleman2003,
bortz2004}. These studies have uncovered the physics of {\sl
underscreening}: a single conduction channel can screen at most $1/2$ unit
of spin, so that at low temperatures the impurity remains magnetic with a
residual spin of $S-1/2$ \cite{mattis1967}. It was found that for $S \geq 1$
the approach to the stable Fermi-liquid fixed point is slow (logarithmic);
this behavior was characterized as that of a singular Fermi liquid
\cite{mehta2005, koller2005}.

The anisotropic Kondo model with $D \neq 0$ (and with additional anisotropy
in the exchange coupling) was studied by the Bethe-Ansatz technique
\cite{schlottmann2000, schlottmann2001}. It was shown that anisotropy can
induce a quantum critical point (i.e. non-Fermi-liquid behavior). The
applicability of the Bethe-Ansatz approach is, however, limited to a set of
models with restrained parameters thus a direct comparison with the model
considered in this work is not possible. Nevertheless, there is qualitative
agreement in that we also find a non-Fermi-liquid state for $D<0$.

The anisotropic Kondo model was also previously studied by the numerical
renormalization group technique in Refs.~\onlinecite{romeike2006} and
\onlinecite{romeike2006b} with the focus on high values of spin $S$ and
easy-axis anisotropy as appropriate for molecular magnets (we note that the
definition of $D$ in the cited works differs in sign from ours). Where
comparisons can be made, our results agree with theirs.

Finally, we mention that related models and physical effects are also
studied in the context of transport spectroscopy of quantum dots and
impurity clusters. The high conductance in the case of underscreening in
spin-$1$ quantum dots and the two-stage Kondo screening were discussed in
Refs.~\onlinecite{posazhenikova2005} and \onlinecite{posazhenikova2007}.
Furthermore, the two-stage Kondo screening is also found in the case of a
single conduction channel in multiple-quantum-dot structures, in particular
in the case of side-coupled quantum dots \cite{vojta2002, cornaglia2005tsk,
sidecoupled, chung2008}. High-spin states can be obtained in systems of
multiple impurities if the conduction-band-mediated exchange interaction is
ferromagnetic \cite{vzporedne, vzporedne2}, which may occur for small
inter-impurity separations. Two-stage Kondo screening and the closely
related ``singlet-triplet'' Kondo effect \cite{izumida2001, pustilnik2001,
pustilnik2001st, sakai2003, pustilnik2003, hofstetter2004} have both been 
intensively studied experimentally \cite{schmid2000, sasaki2000, wiel2002,
kogan2003, fuhrer2004a, granger2005}.

\section{Scaling analysis}
\label{secscaling}

There is a close relation \cite{konik2002restor, schiller2008} between the
Kondo model with the $DS_z^2$ magnetic anisotropy term and the Kondo model
with the XXZ anisotropic exchange constants $J_z$ and $J_\perp$
\cite{anderson1970, tsvelick1983, costi1996akm, costi1998akm, schiller2008}
defined by the Hamiltonian
\begin{equation}
H_\mathrm{K} = J_z s_z S_z + J_\perp \left( s_x S_x +s_y S_y \right).
\end{equation}
To be specific, we consider here the $S=1$ anisotropic Kondo model with both
types of the anisotropy (XXZ exchange and the $DS_z^2$ term, but $E=0$).
Taking into account that the energy of the $S_z=\pm 1$ intermediate states
is higher by $D$ as compared with the energy of the $S_z=0$ state, the
following scaling equations are obtained \cite{anderson1970,
konik2002restor, schiller2008}
\begin{align}
\label{eqJz}
\frac{\dr j_z}{\dr l} &= \frac{1}{1-d} j_\perp^2, \\
\label{eqJp}
\frac{\dr j_\perp}{\dr l} &= \frac{1}{2}\left( 1+\frac{1}{1+d} \right)
j_\perp j_z, \\
\label{eqD}
\frac{\dr d}{\dr l} &= d -\ln2 \left( j_z^2 - j_\perp^2 \right).
\end{align}
Here the scaling parameter is $l=-\ln W$; it runs from 0 to positive
infinity as the energy scale is reduced. Note that here we denote the
half-bandwidth by $W$ (this departs from the NRG convention of denoting it
by $D$). We have introduced dimensionless running coupling constants by
absorbing the density of states $\rho$: $j_\perp=\rho J_\perp$, $j_z=\rho
J_z$, while $d$ is measured in the units of $W$: $d=D/W$. These scaling
equations differ from those in Ref.~\onlinecite{schiller2008} only by the
$d$-dependent factors in Eqs.~\eqref{eqJz} and \eqref{eqJp}, which arise due
to the energy shift by $D$ in the $S_z=\pm 1$ states.

From Eq.~\eqref{eqD} it follows that $d$ will rapidly grow in absolute value
\cite{schiller2008}, since the $d$ term on the right hand side is first
order, while the $j_z^2$ and $j_\perp^2$ terms are second order, hence
smaller. The importance of the axial anisotropy is thus always reinforced at
lower energy scales. Furthermore, starting from initially equal bare
coupling constants, $j_z=j_\perp$, a non-zero $d$ induces XXZ exchange
anisotropy by unequal renormalization of $j_z$ and $j_\perp$, as seen from
Eqs.~\eqref{eqJz} and \eqref{eqJp}. At low energy scales, the physical
properties are qualitatively the same irrespective of the origin of the
anisotropy. From Eq.~\eqref{eqD} we may also anticipate that the $D>0$ case
corresponds to a $J_\perp > J_z$ exchange anisotropic Kondo model, while the
$D<0$ case to a $J_\perp < J_z$ model \cite{schiller2008}.

\section{Numerical results}
\label{secresults}

Calculations were performed using the numerical renormalization group
technique \cite{wilson1975, cragg1979b, cragg1980, krishna1980a, hewson,
bulla2008} which consists of a logarithmic discretization of the continuum
of states of the conduction band electrons, followed by mapping to a
one-dimensional chain Hamiltonian with exponentially decreasing hopping
constants. This Hamiltonian is then diagonalized iteratively by taking into
account one new lattice site at each iteration. We have used discretization
parameter $\Lambda=2.5$ ($\Lambda=2$ for spectral function calculations) and
truncation with the energy cutoff at $E_\mathrm{cutoff}=10\omega_N$, where
$\omega_N$ is the characteristic energy scale at the $N$-th iteration step.
As a further precaution, truncation is always performed in a ``gap'' of
width at least $0.01 \omega_N$, so as not to introduce systematic errors. To
prevent spurious polarization of the residual impurity spin in the $D<0$
case due to floating-point round-off errors, it is helpful to symmetrize the
energies of states which should be exactly degenerate (i.e. the $\pm S_z$
pairs). In all calculations, we have taken explicitly into account the
conservation of charge $Q$; in calculations with only the $DS_z^2$ anisotropy
term ($E=0$), a further conserved quantum number is $S_z$.

In all calculations we set $\rho J=0.1$, therefore the Kondo temperature in
the isotropic case, $\TKO$, is the same for all values of spin $S$
\cite{andrei1983, bortz2004, kaihe2005, vzporedne} and given by
\cite{wilson1975, krishna1980a}:
\begin{equation}
\TKO \approx W \sqrt{\rho J} \exp \left(-\frac{1}{\rho J} \right) 
= 1.4\times 10^{-5} W.
\end{equation}
Extracting $\TKO$ directly from the NRG results we obtain a more accurate
value of $\TKO = 1.16\times 10^{-5} W$. (We use Wilson's definition of the Kondo
temperature \cite{wilson1975, andrei1981, andrei1983}.)

\subsection{Static properties}

We first consider the static (thermodynamic) properties, in particular the
impurity contribution to the magnetic susceptibility, $\chi_\mathrm{imp}$,
and the impurity contribution to the entropy, $S_\mathrm{imp}$. The first
quantity, when multiplied by the temperature, is a measure of the effective
magnetic moment at a given temperature scale, while the second provides
information about the effective impurity degrees of freedom.

\begin{figure*}[htbp!]
\centering
\includegraphics[width=14cm]{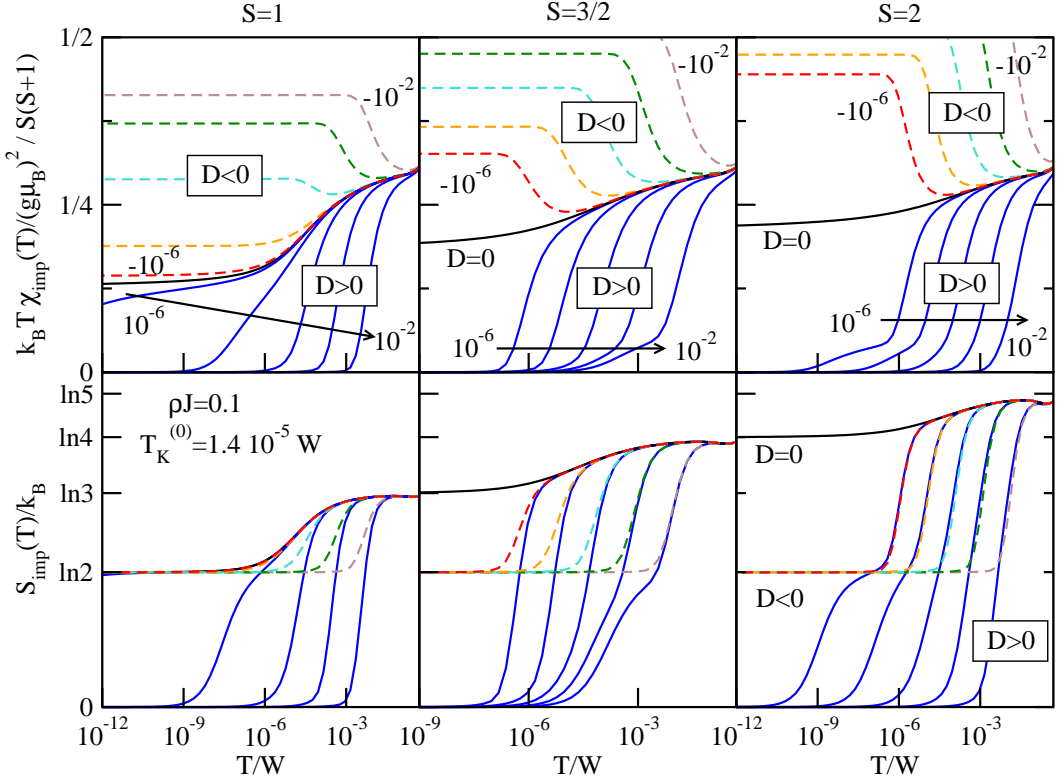}
\caption{(Color online) Thermodynamic properties of the anisotropic Kondo
model. We plot the impurity contribution to the magnetic susceptibility,
$\chi_\mathrm{imp}$, and the impurity contribution to the entropy,
$S_\mathrm{imp}$. Full (blue) lines correspond to positive anisotropy,
$D>0$, dashed (red) lines correspond to negative anisotropy, $D<0$, while
black line is the isotropic case, $D=0$. The arrow indicate in which
direction the anisotropy is increasing. Note that the vertical axis in the
upper panels is additionally rescaled as $1/S(S+1)$.} \label{figb}
\end{figure*}

In Fig.~\ref{figb} we plot the thermodynamic properties of the anisotropic
Kondo model with the axial $DS_z^2$ term ($E=0$) for three different values
of the impurity spin $S$ and for a range of the anisotropy strength. As long
as the temperature is larger than the anisotropy $D$, the curves follow
closely the results for the isotropic case (black curves). At lower
temperatures, the behavior of the system strongly depends on the sign of
$D$. For negative $D$ (easy-axis case), the impurity spin remains partially
unscreened at low temperatures; for small $|D|$, the effective moment is
slightly above the isotropic strong-coupling fixed-point value with spin
$S-1/2$ [i.e. $\mu=(S^2-1/4)/3$] and it saturates with increasing $|D|$ at
the value of $\mu=S^2$. The low-temperature impurity entropy is $\ln 2$ for
all negative $D$. This suggests that the impurity behaves as a residual
two-level system with states $\ket{+}$ and $\ket{-}$, which in the
large-$|D|$ limit become equal to the $\ket{S_z=+S}$ and $\ket{S_z=-S}$
states. The scattering of the low-energy electrons on this residual impurity
states is discussed in App.~\ref{secanalysis}. For spin $S>1$, the effective
moment always increases as the temperature $T$ is decreased below $|D|$,
which reflects the progressive freezing out of the magnetic levels other
than the maximally aligned $S_z=\pm S$ states. For spin $1$, this is still
the case for large enough $|D|$, but for lower $|D|$ the temperature
dependence becomes monotonic. The boundary between the two different
behaviors can be approximately located at $|D| \sim \TKO$. The reason
appears to be that for $|D|<\TKO$, the $S=1$ Kondo effect had already
significantly screened the impurity spin yielding a residual spin $1/2$ by
the time the anisotropy begins to be felt; the only effect of the anisotropy
is then to prevent the screening process from completing, which gives a
residual spin value slightly above $1/2$.

For positive $D$ (hard-axis case), the impurity is non-magnetic at low
temperatures for any value of the impurity spin $S$. Unlike in the easy-axis
case, we see a larger variety of possible behaviors in the approach to the
fixed point depending on the spin and on the anisotropy strength. We discuss
$S=3/2$ and $S=2$ first; these are characteristic representatives for all
half-integer and integer spin cases.

\begin{figure}[htbp]
\centering
\includegraphics[width=7cm]{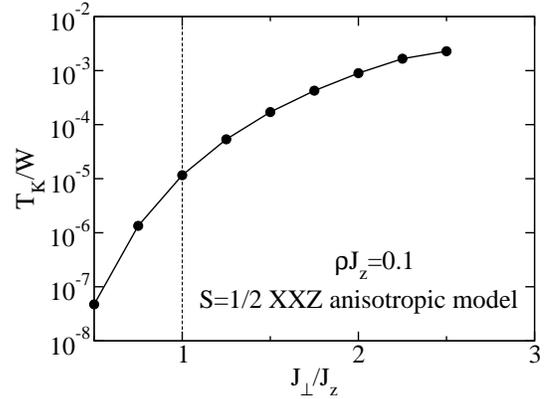}
\caption{The Kondo temperature of the XXZ anisotropic
$S=1/2$ Kondo model as a function of the $J_\perp/J_z$ ratio.}
\label{figtk}
\end{figure}

For half-integer spin and small $D<\TKO$, the curves just follow the results
for the isotropic case until $T \sim D$, when spin states other than the
$S_z=0$ state of the residual (integer!) $S-1/2$ impurity spin freeze out
and the system approaches the non-magnetic ground state exponentially fast.
For sufficiently large $D>\TKO$, the high $|S_z|$ states of the original
impurity spin freeze out exponentially fast at $T \sim D$, this time
yielding a $S_z=\pm 1/2$ doublet which then undergoes spin-$1/2$ Kondo
screening. It must be stressed, however, that after the high-$|S_z|$ states
are frozen out, the effective model in the restrained $S_z=\pm 1/2$ subspace
does {\it not} correspond to an isotropic spin-$1/2$ model, but rather to an
exchange anisotropic spin-$1/2$ model with
\begin{equation}
J_\perp \approx 2J,\quad J_z \approx J.
\end{equation}
The mapping on an anisotropic Kondo model stems from the spin-$3/2$
operators which are given in the matrix notation as
\begin{equation}
\begin{split}
S_x &= \begin{pmatrix}
0 & \frac{\sqrt{3}}{2} & 0 & 0 \\
\frac{\sqrt{3}}{2} & 0 & 1 & 0 \\
0 & 1 & 0 & \frac{\sqrt{3}}{2} \\
0 & 0 & \frac{\sqrt{3}}{2} & 0 
\end{pmatrix}, 
\\
S_z &= \begin{pmatrix}
\frac{3}{2} & 0 & 0 & 0 \\
0 & \frac{1}{2} & 0 & 0 \\
0 & 0 & -\frac{1}{2} & 0 \\
0 & 0 & 0 & -\frac{3}{2} \\
\end{pmatrix}.
\end{split}
\end{equation}
In the $S_z=\pm 1/2$ subspace, $S_z$ yields $\frac{1}{2} \sigma_z$, while
$S_{x,y}$ yield $2 \times \frac{1}{2} \sigma_{x,y}$, i.e. twice the
spin-$1/2$ operators in the transverse directions. The Kondo
temperature is thus given by the expression for the XXZ anisotropic
Kondo model:
\begin{equation}
\begin{split}
T_K &\approx W \exp \left( -\frac{\alpha}{\rho J_z} \right), \\
\alpha &= \frac{\arctan \gamma}{\gamma}, \\
\gamma &= \sqrt{ \left(\frac{J_\perp}{J_z}\right)^2-1 },
\quad \text{for}\, J_\perp \geq J_z,
\end{split}
\end{equation}
which can be derived from the scaling equations to second order
\cite{romeike2006}. For reference we also note that the expression for
$J_\perp < J_z$ can be obtained by analytic continuation, giving
\begin{equation}
\begin{split}
T_K &\approx W \exp \left(-\frac{\alpha}{\rho J_z} \right), \\
\alpha &= \frac{\mathrm{arctanh} \gamma}{\gamma}, \\
\gamma &= \sqrt{ 1-\left(\frac{J_\perp}{J_z}\right)^2 },
\quad \text{for}\, J_\perp \leq J_z.
\end{split}
\end{equation}
These expressions somewhat overestimate the true Kondo temperature since
they do not include the $\sqrt{\rho J_z}$ factors; numerically calculated
Kondo temperatures are shown in Fig.~\ref{figtk}.

The value of the Kondo temperature of the effective spin-$1/2$ screening
process depends on $D$ and it can exceed $\TKO$ since $\alpha < 1$ in the
transverse dominated $J_\perp/J_z>0$ case; in the infinite-$D$ limit,
$T_K(S=1/2)$ goes to 
\begin{equation}
T_K(S=1/2) \approx 9\times 10^{-4} W.
\end{equation} 
This result fully agrees with $T_K$ in the exchange-anisotropic Kondo model,
Fig.~\ref{figtk}. For finite $D$, the effective ratio $J_\perp/J_z$ may be
different from 2, since there is a temperature range where the $S_z=\pm 3/2$
levels still affect the renormalization of the exchange interaction before
they completely freeze out. The intrinsic anisotropy of the effective
spin-$1/2$ Kondo model also plays a role in the splitting (shifting) of the
Kondo resonance in a magnetic field: for strong magnetic field $B_z$ in the
$z$-direction, we expect a shift of $\sim g \mu_B B_z$, and a shift
approximately twice as large for fields in the transverse directions
\cite{heinrichprivate}. Finally, we note that for larger spin values we have
$J_\perp=3 J$ (spin $5/2$), $J_\perp=4J$ (spin $7/2$), and in general 
\begin{equation}
J_\perp = (S+1/2)J.
\end{equation}

\begin{figure}[htbp]
\centering
\includegraphics[width=7cm]{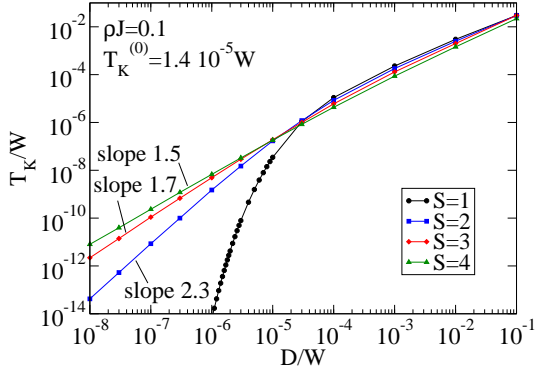}
\caption{(Color online) The cross-over temperature (indentified with the spin-$1/2$ Kondo
temperature $T_K$ for $D < \TKO$) in integer-spin anisotropic Kondo models
for the hard-axis case.}
\label{figh}
\end{figure}

For integer $S \geq 2$, the situation is the opposite. Now the regime of the
exponential approach to the $S_z=0$ ground state corresponds to $D > \TKO$.
For $D < \TKO$, on the other hand, the low-temperature behavior is again
that of the spin-$1/2$ Kondo model. This is made possible by the combined
effect of the spin-$S$ Kondo screening at higher temperatures giving rise to
the $S-1/2$ (i.e. half-integer) residual impurity spin, and of the axial
anisotropy which leads to freezing out of the high-spin $S_z=\pm
(S-1/2),{\ldots} , \pm 3/2$ levels of this residual object to finally give
rise to a residual spin-$1/2$ object. This residual spin-$1/2$ then finally
undergoes a spin-$1/2$ Kondo screening. At first sight it might even seem
surprising that the residual spin is compensated at all, given that in the
isotropic high-spin Kondo model the residual exchange interaction is
ferromagnetic, yet for $D>0$ the induced exchange anisotropy (see
Sec.~\ref{secscaling}) is of the type which leads to complete screening of
the impurity spin \cite{konik2002restor}.  Since the residual $(S-1/2)$ spin
is an extended object, there is no simple mapping on the effective
spin-$1/2$ Kondo model and it appears difficult to estimate the effective
exchange constants $J_\perp$ and $J_z$ even in the $D \to 0$ limit. The
effective bandwidth on the other hand is clearly given by $W_\mathrm{eff} =
\alpha D$, where $\alpha$ is some constant of order 1. The Kondo temperature
is thus given by $T_K(S=1/2) \approx D f(D)$, where $f(D)$ turns out to be a
power-law function with non-integer exponent, $f(D) \sim D^{1.3}$ for $S=2$
and $f(D) \sim D^{0.7}$ for $S=3$, see Fig.~\ref{figh}. For large spin the
Kondo temperature is simply proportional to $D$.

We would like to emphasize the important fact that the original impurity
spin $S$ and the residual spin $S-1/2$ after the Kondo screening belong to
different classes (one is integer, the other half-integer). For this reason
we always find different behavior depending on which of the $D$ and $\TKO$
is smaller. Furthermore, this explains similarities between integer $S$ for
$D<\TKO$ and half-integer $S$ for $D>\TKO$.

Finally we consider the special case of integer spin 1. For large $D >
\TKO$, the behavior is the same as for other integer spins. For small $D <
\TKO$, the approach to the strong-coupling fixed point is of the spin-$1/2$
Kondo model type. However, we find that the second Kondo temperature $\TKt$
depends exponentially on the $D/\TKO$ ratio, see Fig.~\ref{figh}, a
situation strongly reminiscent of the two-stage Kondo screening
\cite{jayaprakash1981} in the side-coupled impurity systems \cite{vojta2002,
cornaglia2005tsk, sidecoupled, peters2006fluc, chung2008}. At $\TKO$ the
spin is screened from $S=1$ to $S=1/2$ by the spin-$1$ Kondo effect, then at
$\TKt$ from $S=1/2$ to $S=0$ by the spin-$1/2$ Kondo effect. It is possible
to fit the results with
\begin{equation}
\TKt = \alpha \TKO e^{-\gamma \frac{\TKO}{J_\mathrm{eff}}}
\end{equation}
where
\begin{equation}
J_\mathrm{eff} = D + c \TKO
\end{equation}
with coefficients $\alpha=0.052$, $\gamma=2.531$, $c=0.0661$. This fit is
only of phenomenological value: the situation here is different from the one
in the side-coupled impurity case, where it is possible to interpret the
results in terms of screening of the second impurity by the quasiparticles
resulting from the first stage of the Kondo screening and where
$J_\mathrm{eff}$ corresponds to real exchange coupling.

\begin{figure}
\centering
\includegraphics[width=7cm]{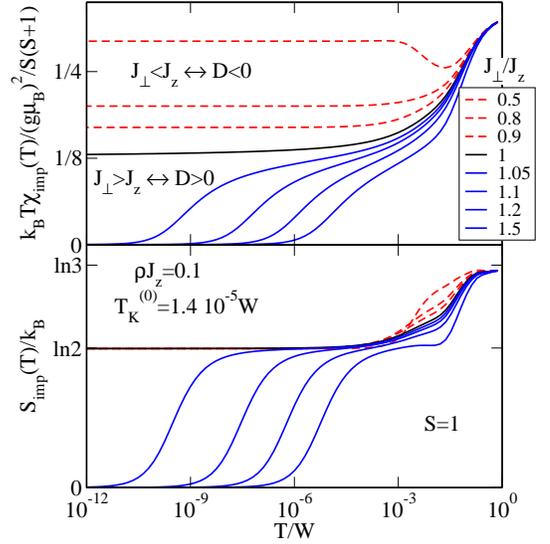}
\caption{(Color online) Thermodynamic properties of the $S=1$ Kondo
model with XXZ exchange anisotropy.}
\label{figd}
\end{figure}

In Fig.~\ref{figd} we plot the thermodynamic quantities for the related 
$S=1$ Kondo model with XXZ exchange anisotropy \cite{schiller2008}. As
expected, we find that $J_\perp < J_z$ models behave similarly to $D<0$, and
$J_\perp > J_z$ similarly to $D>0$. In the presence of both symmetries, the
two anisotropies may either enhance each other or compete.

\begin{figure*}
\centering
\includegraphics[clip,width=14cm]{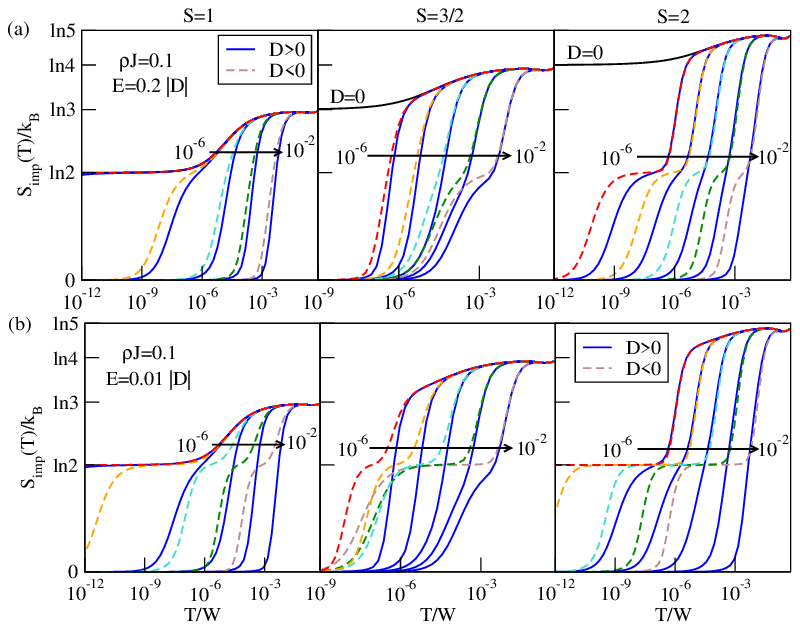}
\caption{(Color online) Impurity contribution to the entropy of the
anisotropic Kondo model with $D S_z^2 + E(S_x^2-S_y^2)$ terms with constant
ratio (a) $E/|D|=0.2$ and (b) $E/|D|=0.01$.  We plot
$D=10^{-6},10^{-5},{\ldots} ,10^{-2}$; the arrow indicate in which direction
the anisotropy is increasing. }
\label{figb3}
\end{figure*}

In Fig.~\ref{figb3} we plot the impurity contribution to the entropy for the
fully anisotropic problem with the $DS_z^2$ and $E(S_x^2-S_y^2)$ terms
(impurity contribution to the magnetic susceptibility cannot be easily
computed since $S_z$ is no longer a good quantum number). We consider both
the case when $E$ is large [subfigure a) with $E=0.2|D|$; such ratio is
found, for example, for integer spin-$2$ Fe impurities on CuN/Cu surfaces
where $D=\unit[-1.5]{meV}$ and $E=\unit[0.3]{meV}$, and for half-integer
spin-$5/2$ Mn impurities on the same surface \cite{hirjibehedin2007}] and
for the case where $E$ is much lower than $|D|$ [subfigure b) with
$E=0.01|D|$; no such magnetic adatom/surface system had been identified so
far in the negative-$D$ systems, but $E$ appears to be much smaller than $D$
in the positive-$D$ Co on CuN/Cu system].

\begin{figure}[htbp]
\centering
\includegraphics[width=7cm]{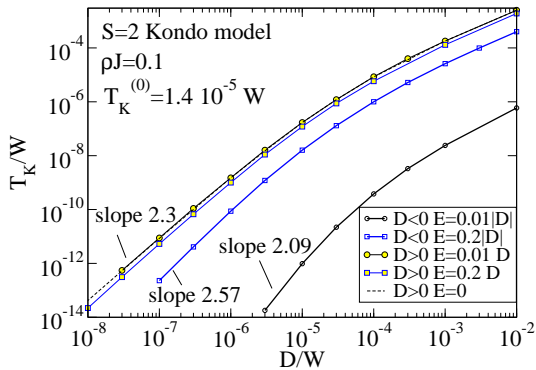}
\caption{(Color online) The cross-over temperature (indentified with the spin-$1/2$ Kondo
temperature $T_K$ for $D < \TKO$) in the $S=2$ anisotropic Kondo model
with transverse anisotropy.}
\label{fighE}
\end{figure}

The ground state for $E \neq 0$ is non-degenerate for all parameters. For
$D>0$, the $S_x^2-S_y^2$ operator is an irrelevant perturbation and the
thermodynamic behavior of the system is hardly affected. For $D<0$, however,
the perturbation is relevant and drives the system to a different,
non-magnetic ground state for all $S$. For $S=1$, the degeneracy is lifted
on the scale of $E$ if $E>\TKO$; for $E<\TKO$ there appears to be Kondo-like
screening of the residual spin-$1/2$ with exponential reduction of the
second Kondo temperature, similar to what happens in the $D>0$ case. Thus
the $E$-term can induce two-stage Kondo behavior even in the $D<0$ case. For
half-spin $S=3/2$, the degeneracy is again lifted on the scale of $E$ if
$E<\TKO$. If $E>\TKO$, previously known Kondo effect with pseudo-spin $1/2$
occurs \cite{romeike2006}; the Kondo temperature depends on parameters in a
non-trivial way. For integer spin $S=2$, we find degeneracy lifting on the
scale of $E$ if $E>\TKO$, and a pseudo-spin $1/2$ Kondo effect if $E<\TKO$.
The effective bandwidth is now given by $W_\mathrm{eff}=\alpha E$, where
$\alpha$ is some constant of order 1, so the Kondo temperature is given by
$T_K \approx E f(D,E)$ where $f(D,E)$ has power-law behavior as a function
of $D$ with a $D/E$ (and spin $S$) dependent exponent, see Fig.~\ref{fighE}.

We conclude that the easy-axis systems with transverse anisotropy behave
rather similarly to hard-axis systems; in the conditions for the emergence
of the pseudo-spin $S=1/2$ Kondo effect the quantity $E$ takes the place of
$D$.

\subsection{Properties of the $D<0$ systems}

For any $S \geq 1$ and $D<0$, $E=0$, the ground state is twofold degenerate.
The fixed-point spectra depend on the value of $D$, thus the $S_z^2$ term is
a marginal operator. At low temperatures, these systems have a Curie-like
magnetic response with fractional spin
\begin{equation}
\chi_\mathrm{imp}(T \to 0) = (g\mu_B)^2 \frac{\mathcal{C}(D)}{k_B T}.
\end{equation}
This is reminiscent of the fractional-spin non-Fermi-liquid fixed points in
the pseudo-gap Kondo model with $\rho(\omega) \sim |\omega|^r$ density of
states for spin 1 and $r>0$ \cite{gonzalez1998, florens2005, withoff1990,
chen1995pseudogap} and in the related power-law Kondo model in the case of
ferromagnetic coupling and $r<0$ \cite{vojta2002eurpjb}. There are
nevertheless some notable differences. The fixed points in the pseudo-gap
Kondo model are found to be unstable with respect to the particle-hole
symmetry breaking. The fixed points we find are, however, stable with
respect to particle-hole symmetry breaking: both anisotropy, $S_z^2$, and
potential scattering, $n_{f_0}$, are marginal operators. Furthermore, the
fixed point in the anisotropic Kondo model has entropy $\ln 2$ (i.e.
impurity behaves as a two-level system), while the fixed point in the
pseudo-gap Kondo model has entropy $(1+2r)\ln 2$ \cite{gonzalez1998,
florens2005}.

\begin{figure}[htbp]
\centering
\includegraphics[width=6cm]{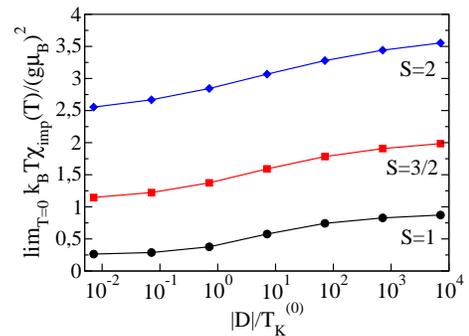}
\caption{(Color online) Curie constant as a function of the $|D|/T_K^{(0)}$ ratio
for $S=1, 3/2, 2$ Kondo models with easy-axis anisotropy, $D<0$.}
\label{fige}
\end{figure}

In Fig.~\ref{fige} we plot the Curie constant as a function of the $D/\TKO$
ratio. The transition from the isotropic limit
\begin{equation}
\mathcal{C} = (S-1/2)(S+1/2)/3=(S^2-1/4)/3,
\end{equation}
to the saturated easy-axis anisotropic behavior, 
\begin{equation}
\mathcal{C} = S^2,
\end{equation}
is approximately logarithmic, i.e. very slow. This reflects the underlying
underscreened Kondo effect and the partial screening of the impurity moment
which also feature similarly slow logarithmic dependence of the magnetic
susceptibility on temperature and magnetic field.

\begin{figure}[htbp]
\centering
\includegraphics[width=8cm]{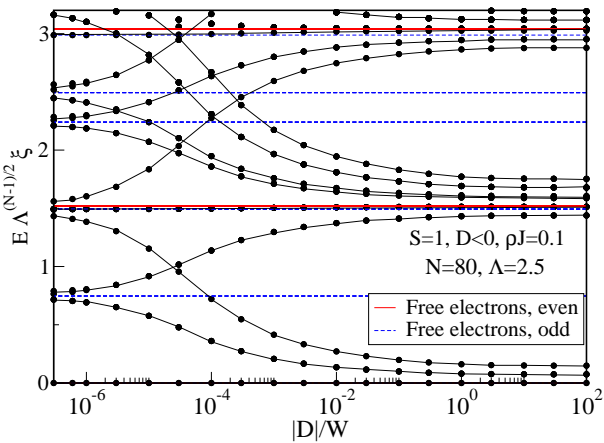}
\caption{(Color online) Energy levels at iteration $N=80$ ($T \sim 10^{-16}W$)
as a function of the anisotropy $D$ in the spin-1 Kondo model
with easy-axis anisotropy, $D<0$.}
\label{figg}
\end{figure}

In Fig.~\ref{figg} we plot the energy levels of the spin-1 negative-$D$
model as a function of the anisotropy $D$. For small $|D|$, the levels
approach the free electron spectrum with $\pi/2$ phase shift (which is due
to the Kondo effect) with additional two-fold degeneracy of all levels due
to the decoupled spin-$1/2$ residual impurity spin. For large $|D|$, the
levels tend towards a spectrum with states clustered near the energies of
the free electron spectrum with zero phase shift (no Kondo effect) with
additional splitting due to residual anisotropic exchange coupling that will
be studied in App.~\ref{secanalysis}. For intermediate $|D|$, the spectrum
is more complex. It can be characterized as free electron spectrum with some
intermediate phase shift with additional splitting due to residual exchange
interaction; as $|D|$ is swept from $0$ to $+\infty$, the (spin-dependent)
phase shift is reduced from $\pi/2$ to 0, while the nature of the residual
impurity spin changes from isotropic spin-$1/2$ doublet to the anisotropic
$S_z=\pm 1$ magnetic doublet, and the residual exchange constants increase
from 0 to some finite value of the order of the bare exchange coupling $J$,
see also App.~\ref{secanalysis}.

\subsection{Dynamic properties}

To characterize dynamic properties of the anisotropic models, we calculate
the T-matrix and the dynamical spin susceptibility \cite{costi1994,
hofstetter2000, peters2006, weichselbaum2007}. The T-matrix for the Kondo
model can be determined by computing the Green's function \cite{costi2000,
zarand2004, koller2005}
\begin{equation}
G(\omega) = \correl{S^- f_{0\downarrow} + S^z f_{0\uparrow};
S^+ f^\dag_{0\downarrow} + S^z f^\dag_{0\uparrow}},
\end{equation}
where $S^{\pm} = S_x \pm i S_y$ are the impurity spin operators and
$f^\dag_{0\sigma}$ creates an electron with spin $\sigma$ on the first site
of the hopping Hamiltonian \cite{wilson1975, bulla2008}. Assuming constant
exchange constant $J_{\vc{k},\vc{k'}}\equiv J$, the T-matrix is then given
by
\begin{equation}
T(\omega) = J^2 G(\omega).
\end{equation}
This quantity contains information on both elastic and inelastic scattering
rate (cross section) \cite{zarand2004, koller2005, borda2007inelastic}:
\begin{equation}
\begin{split}
\sigma_\mathrm{total}(\omega)/\sigma_0 = \rho \pi^2 \left(
-\frac{1}{\pi} \mathrm{Im} T(\omega) \right), \\
\sigma_\mathrm{el}(\omega)/\sigma_0 = \rho^2 \pi^2 |T(\omega)|^2,\\
\sigma_\mathrm{inel}(\omega) = \sigma_\mathrm{total}(\omega)
-\sigma_\mathrm{el}(\omega),
\end{split}
\end{equation}
where $\sigma_0$ is the scattering rate in the case of unitary scattering
\cite{koller2005}. These scattering rates can, in turn, be related to the
amplitude of the impurity-related spectral features in scanning tunneling
spectroscopy experiments \cite{madhavan1998, li1998, vojta2002stm,
cornaglia2003stm}, thus in the following we will call the quantity
$\sigma_\mathrm{total}(\omega)$ the ``conductance'' and we will express it
in the units of $\sigma_0$. We also note that the zero-temperature
scattering rate in a Fermi-liquid system is
\begin{equation}
\sigma_\mathrm{total}(\omega \to 0) = \sigma_\mathrm{el}(\omega \to 0)
= \sigma_0 \sin^2(\delta),
\end{equation}
where $\delta$ is the quasiparticle scattering phase shift, while
$\sigma_\mathrm{inel}(\omega \to 0)=0$.

The imaginary part of the dynamical spin susceptibility
$\cchi_{\alpha\alpha}(\omega)$ (with $\alpha=x,y,z$) is defined as
\begin{equation}
\cchi_{\alpha\alpha}(\omega) = -\frac{1}{\pi} \mathrm{Im}
\correl{S_\alpha;S_\alpha}.
\end{equation}
The dynamical spin susceptibility is in principle observable in
tunneling-current-noise spectroscopy using spin-polarized STM
\cite{balatsky2002esrstm, nussinov2003}. Our results thus provide
information on how the Kondo effect in the presence of magnetic anisotropy
modifies the current noise. In addition, the dynamical spin susceptibility
contains information on the differential cross section
$\sigma(\omega,\Delta)$, where $\Delta$ is the energy exchange
\cite{garst2005}; the differential cross sections is also possibly measurable
\cite{garst2005}.

\begin{figure*}
\centering
\includegraphics[width=12cm]{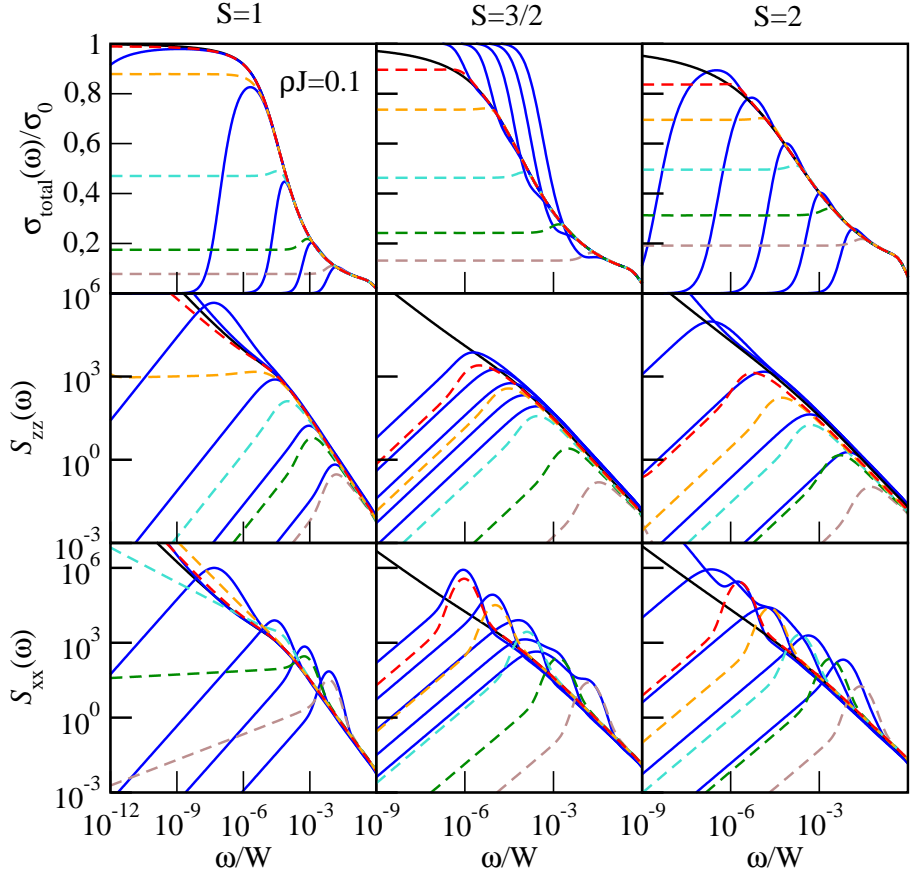}
\caption{(Color online) Dynamic properties of the anisotropic Kondo model.
We plot the spectral function of the $T$-matrix (``conductance'') and the
longitudinal and transversal dynamical suspectibility
$\cchi_\mathrm{zz}(\omega)$ and $\cchi_\mathrm{xx}(\omega)$. The curves
correspond to $|D|=10^{-2}, \ldots, 10^{-6}$; the curve styles are the same
as in Fig.~\ref{figb}. Note that for $S=2$ and $D=10^{-6}$ the susceptibility 
curves do not diverge; they start decreasing at an energy scale outside the
displayed range.
}
\label{figc}
\end{figure*}

In Fig.~\ref{figc} we plot the dynamic properties of the anisotropic Kondo
model with the $DS_z^2$ term ($E=0$). We discuss first the conductance. In
the isotropic $D=0$ case, the conductance always rises to the unitary limit,
albeit the approach to this limit is slow (logarithmic) \cite{coleman2003,
mehta2005, koller2005, posazhenikova2005}. Unitary scattering can be
explained by the well known fact that in the isotropic case a single
conduction channel screens precisely 1/2 unit of the impurity spin and, in
doing so, the low-energy conduction band electrons gather a $\pi/2$
scattering phase shift. 

In the hard-axis $D>0$ case, the conductance at low temperatures is zero for
all integer values of the impurity spin, while it is unitary for all
half-integer spins. For $S=1$ and $D<\TKO$, this can be explained by the
two-stage Kondo screening: the first $S=1$ screening stage leads to
increased conductance as in the isotropic case, however the conductance
drops to zero at a lower temperature scale $\TKt$ when the residual
spin-$1/2$ is compensated in the second screening stage. This is similar to
the screening of a $S=1$ impurity by two conduction channels with unequal
exchange constants \cite{nozieres1980, sasaki2000}, but here the two-stage
screening occurs with a single channel. Non-monotonous energy-dependence of
the spectral function is also found in the case of side-coupled impurities,
but there both screening stages are of the $S=1/2$ kind \cite{vojta2002,
cornaglia2005tsk, sidecoupled, chung2008}. For large $D > \TKO$, the
anisotropy makes the impurity non-magnetic, there is no exchange scattering
at energy scales below $D$ and the low-temperature conductance drops to
zero. Note that in both cases the stable fixed point is the same, the ratio
$D/\TKO$ merely determines by which mechanism the $S=1$ Kondo screening is
interrupted by the anisotropy; the transition between the two regimes is
smooth.

For higher integer spins (represented in Fig.~\ref{figc} by the $S=2$ case),
the situation is in some respect similar. For $D<\TKO$, the conductance at
first increases during the initial spin-$S$ Kondo screening, there is even a
slight bump above the result for the isotropic case on the energy scale of
$D$ when the effective spin doublet is formed and there is additional
resonantly enhanced scattering. Spin-$S$ Kondo screening yields a
half-integer effective impurity spin, thus a spin-$1/2$ doublet on the
energy scale $D$. The spin-$1/2$ Kondo screening then leads to a decrease in
conductance to zero. As previously discussed, the main difference from the
$S=1$ case is that there is no exponential reduction of the energy scale for
$D<\TKO$.

For half-integer spins (represented in Fig.~\ref{figc} by the $S=3/2$ case),
the conductance at low temperatures is unitary, as in the isotropic case.
For $D>\TKO$, this may be explained by the fact that the anisotropy leads in
this case to a low-laying doublet formed by the $S_z=\pm 1/2$ magnetic
doublet which undergoes Kondo screening like in the conventional spin-$1/2$
Kondo model. The doublet is formed before the spin-$S$ Kondo screening
commences and the mapping to the spin-$1/2$ Kondo model is a good
approximation: the Kondo resonance then has approximately Lorentzian form.
It may be noted that there is again an additional peak in $\sigma(\omega)$
at $\omega \sim D$ due to resonantly enhanced scattering by the $S_z \to S_z
\pm 1$ processes. For $D<\TKO$, the conductance curve first follows the
logarithmically increasing isotropic spin-$S$ curve, then at $D$ the
approach to the unitary limit becomes faster.

We emphasize the marked difference between the integer spin $S=2$ and the
half-integer spin $S=3/2$ regarding the energy scale where the limiting
value of $\sigma_\mathrm{total}$ ($0$ viz. $1$) is approached: for $S=2$,
the characteristic frequency decreases faster than linearly with $D$ for
small $D$, while for $S=3/2$ it tends to increase very slowly with $D$ for
large $D$. In both cases this behavior is due to the underlying effective
$S=1/2$ Kondo physics; for $S=2$ it implies that the effective exchange
constant becomes smaller as $D$ is reduced, while for $S=3/2$ this is due to
the saturation of the effective $J_\perp/J_z$ ratio for large $D$ as already
discussed.

\begin{figure}[htbp]
\centering
\includegraphics[width=8cm]{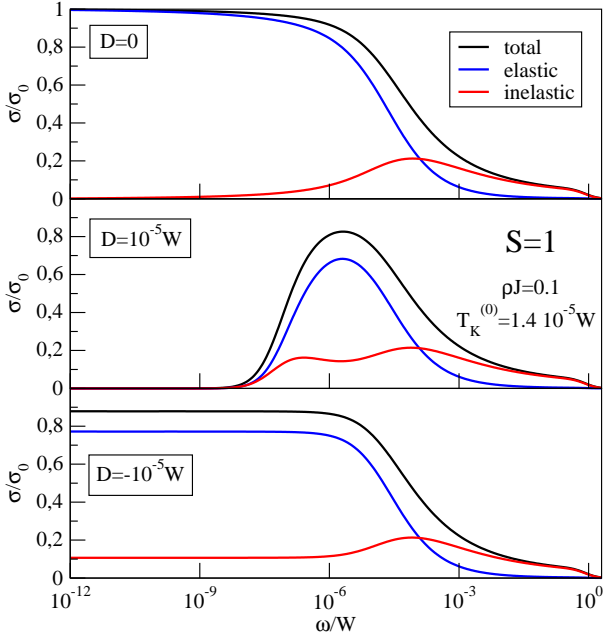}
\caption{(Color online) Total, elastic and inelastic scattering cross sections for
the $S=1$ Kondo model with no anisotropy, positive anisotropy and
negative anisotropy.}
\label{figcc}
\end{figure}

We now consider the easy-axis $D<0$ case. For all $S$, the conductance is
found to saturate at some $D$-dependent finite value here. This reflects the
presence of a residual uncompensated impurity spin which induces scattering
at low temperatures which is neither unitary nor zero (these are the only
two possibilities compatible with a Fermi-liquid system in the presence of
the particle-hole symmetry which imposes the restriction $\delta \equiv
-\delta \mod \pi$ on the quasiparticle scattering phase shift $\delta$, thus
$\delta=0$ or $\delta=\pi$). This implies that the stable fixed point may be
classified as a non-Fermi liquid, which is confirmed by the non-vanishing
inelastic scattering rate at the Fermi level, see Fig.~\ref{figcc}. In the
limit $|D|\to 0$, this non-Fermi-liquid fixed point evolves continuously
into a singular Fermi-liquid fixed point of the isotropic underscreened
Kondo model. Since the approach to this limit is slow (in particular for $S
\geq 3/2$), even very small negative axial anisotropy will lead to
significantly reduced conductance at low temperatures. In the following we
show that this behavior is modified by non-zero $E$ (see also
Ref.~\onlinecite{romeike2006}). Nevertheless, if $E \ll |D|$, there might
exist a temperature range where the non-Fermi-liquid behaviour is
observable.

We now consider the imaginary part of the dynamic spin susceptibility
$\cchi_{\alpha\alpha}(\omega)$. We recall that these quantities are given at
zero temperature by
\begin{equation} 
\cchi_{\alpha\alpha}(\omega) = -\frac{1}{\pi} \frac{1}{Z_N} 
\sum_{i,j_\mathrm{GS}} |\langle i
| S_\alpha | j_\mathrm{GS} \rangle |^2 \delta\left( \omega-(E_i-E_{\mathrm{GS}})
\right)
\end{equation}
where the index $i$ runs over the excited states and the index
$j_{\mathrm{GS}}$ over the degeneracy of the ground state, while $Z_N$ is
the zero-temperature partition function (equal to the degeneracy of the
ground state). In the absence of the coupling to the conduction band, the
transverse susceptibility $\cchi_{xx}(\omega)$ features delta peak(s) at
\begin{equation}
\omega=|DS_z^2 - D(S_z-1)^2| = |D(2S_z-1)|,
\end{equation}
since $S_x$ couples neighboring levels. For $S=1$, there is thus a peak at
$|D|$, for $S=3/2$ a peak at $2|D|$ and for $S=2$, depending on the sign of
$D$, a peak at $D$ (for $D>0$) or at $3|D|$ (for $D<0$). The longitudinal
susceptibility $\cchi_{zz}(\omega)$ of a decoupled impurity is zero.

The $S=1$ case has some special features, so we focus first on the regular
cases, $S \geq 3/2$. The longitudinal susceptibility $\cchi_{zz}(\omega)$
always has a single peak. In the hard-axis $D>0$ case we find that the peak
occurs at the energy scale of the effective $S=1/2$ Kondo effect (when it
occurs, i.e. $D>\TKO$ for $S=3/2$ and $D<\TKO$ for $S=2$, respectively) or
at $\sim D$ in the regime with no $S=1/2$ Kondo effect. In both cases this
corresponds to the energy scale where the conductance approaches its
limiting value, as discussed above. This result is expected, since in the
presence of the Kondo effect the largest magnetic fluctuations always occur
on the energy scale of $T_K$, which is a reflection of the anomalously
strong spin-flip scattering of the conduction band electrons off the
impurity; in the absence of the Kondo effect, magnetic fluctuations occur on
the scale of the local magnetic excitation energy, in this case the
anisotropy $D$. In the easy-axis $D<0$ case, the position of the peak in
$\cchi_{zz}(\omega)$ is always $\sim|D|$.

For all $D$ and all $S \geq 3/2$, the approach to the zero frequency limit
is linear, while the behavior at high frequencies is described approximately
by a power law $\cchi(\omega) \sim \omega^{-\gamma}$ with exponents $\gamma$
that are slightly above 1 \cite{costi1998akm}. 

The transverse susceptibility $\cchi_{xx}(\omega)$ is more complex. In the
hard-axis $D>0$ case, we find a two-peak behavior in the parameter regime
with $S=1/2$ Kondo effect: the first peak corresponds to the scale of $D$,
the second to the scale of $T_K$. In the transition regime $D \sim \TKO$,
the two peaks merge into a single peak which then follows the scale of
anisotropy $D$. In the easy-axis $D<0$ case, the curves again have a single
peak at $\sim |D|$. At this point a comment on the peak width is in order:
the peaks at $\sim |D|$ are over-broadened due to the broadening procedure
used in the NRG method, thus the very narrow peaks take the form of the
broadening kernel \cite{garst2005}.

We now finally turn to $S=1$. For hard-axis $D>0$ anisotropy, we find for
both $\cchi_{zz}(\omega)$ and $\cchi_{xx}(\omega)$ a single peak centered at
$\omega \sim D$ (for $D>\TKO$) or at $\omega \sim T_K^{(2)}$ (for $D<\TKO$).
In $\cchi_{xx}(\omega)$, there is no additional peak on the scale of $D$ for
$D < \TKO$, as was the case for the spin-$2$ model, but we observe a change
of slope at $\omega \sim \TKO$. Even more peculiar are the results for the
easy-axis $D<0$ case. The longitudinal susceptibility has a linear frequency
dependence for $|D| \gtrsim \TKO$, but for $|D| \lesssim \TKO$ we observe
$\omega^\alpha$ scaling with exponent $\alpha$ which depends on the
anisotropy $D$ and which turns negative approximately at $|D| \sim \TKO$,
i.e. the susceptibility becomes divergent. In the transverse susceptibility
we find a strong deviation from linear scaling for all values of $|D|$, with
divergent behavior already for $|D|$ much larger than $\TKO$. We emphasize
that for all spin values $S$, the easy-axis $D<0$ systems exhibit
non-Fermi-liquid features (such as finite inelastic scattering at
$\omega=0$), but only for $S=1$ is the dynamic spin susceptibility divergent
at low frequencies. 

We note that a linear frequency dependence of $\cchi_{\alpha\alpha}(\omega)$
at low frequencies must hold in Fermi-liquid systems as mandated by the
Korringa-Shiba relations \cite{shiba1975, costi1996akm, hofstetter2002akm,
garst2005}. For $D<0$ the system is, however, non-Fermi liquid and the
static magnetic susceptibility is diverging for all $S$, thus
$\cchi(\omega)$ is not expected to be linear. We view the fact that it is
linear for $S \geq 3/2$ merely as a coincidence.

We finish our discussion of dynamic properties by showing that in the
presence of the transverse anisotropy term $E(S_x^2-S_y^2)$, all
non-Fermi-liquid features disappear, Fig.~\ref{figcc}. In particular, we
observe that for all integer $S$ the low-temperature conductance is zero,
while for all half-integer $S$ it is unitary, irrespective of the sign and
size of the axial anisotropy $D$. Furthermore, the spin susceptibility
always has linear frequency dependence at low frequencies. This agrees with
the results for static properties.

\begin{figure*}
\centering
\includegraphics[width=12cm]{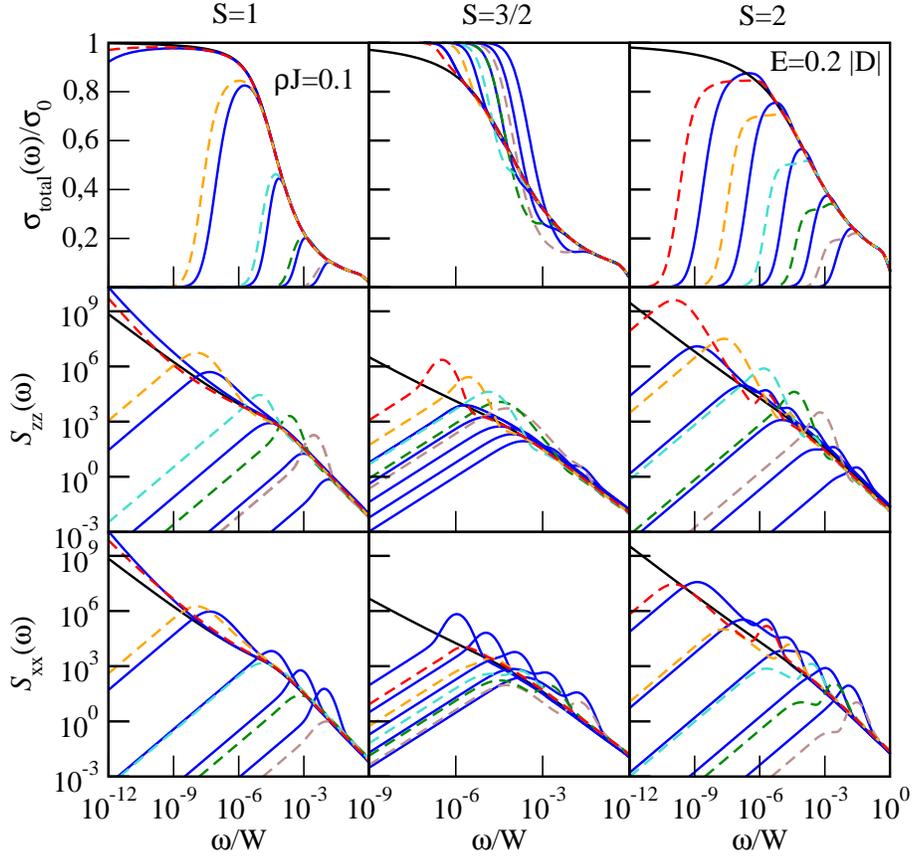}
\caption{(Color online) Dynamic properties of the anisotropic Kondo model
with the $E(S_x^2-S_y^2)$ term. $E=0.2|D|$. The curve styles
are the same as in Fig.~\ref{figb}.
}
\label{figc3}
\end{figure*}

\section{Conclusion}

Due to the presence of the magnetic anisotropy, the impurity spin is always
completely compensated at low enough temperature, even in the case of
high-spin impurities coupled to a single continuum channel. This goes
against the common wisdom that a single channel can screen at most $1/2$
unit of spin, which results in a residual spin of $S-k/2$ in the case of $k$
screening channels: such a behavior will only be observable for isotropic
systems, a situation which is rather unlikely for magnetic impurities
adsorbed on surfaces.

The approach to the fully screened impurity spin with decreasing energy
scale depends on the impurity spin $S$, the sign of axial anisotropy $D$,
and on the hierarchy of the energy scales, in particular on the dichotomy $D
< \TKO$ or $D > \TKO$ in the positive-$D$ models. For positive $D$
(hard-axis case), we find spin-$1/2$ Kondo screening at low temperatures in
the weak-anisotropy case $D < \TKO$ for integer spin (with exponentially
reduced Kondo temperature in the special case of spin 1) and in the
strong-anisotropy case $D > \TKO$ for half-integer spin; in other cases, the
effective impurity moment drops to zero exponentially on the energy scale of
$D$. The transverse anisotropy $E$ does not play an important role for
$D>0$. For negative $D$ (easy-axis case), however, it changes the behavior
of the system completely. For $E=0$, the impurity spin would be only
partially screened, leading to residual spin-dependent scattering and
non-Fermi-liquid behavior. In real systems $E \neq 0$ and the behavior of
$D<0$ becomes similar to that of the $D>0$ systems: for $S=1$ we again find
two-stage Kondo screening for $E < \TKO$, for $S=3/2$ there is effective
spin-$1/2$ Kondo effect for $E > \TKO$, while for $S=2$ we find effective
spin-$1/2$ screening for $E < \TKO$.  If $E \ll |D|$, some features of the
non-Fermi-liquid behavior might be observable in intermediate temperature
respectively energy regimes. The zero-temperature conductance is found to be
zero for all integer spins and unitary for all half-integer spins for both
signs of $D$ (assuming $E \neq 0$). For integer spin and $E, D < \TKO$, the
temperature dependence of the conductance is expected to be non-monotonous
with increased conductance in the $T_K(S=1/2) < T < \TKO$ range. The most
likely candidate for observation of underscreening is the $S=1$ case due to
the exponential reduction of the second Kondo temperature, which is a
special feature of the $S=1$ model and is not found for other integer spins.

We conclude by noting that taking the magnetic anisotropy effects into
account is essential in interpreting scanning tunneling spectra of magnetic
impurities measured at low temperatures ($T \ll |D|$). The appropriate
low-temperature effective impurity model depends on the parameters of the
original physical model. While most interpretations of the surface Kondo
effect have been based on the Anderson impurity model (which maps on the
isotropic Kondo model), it is more likely that the appropriate effective
model is, in fact, some anisotropic Kondo model. This is especially true for
impurities on decoupling layers, but also applies to impurities adsorbed
directly on metallic surface where due to stronger hybridisation charge
fluctuation effects also play a role. Appropriate description in terms of an
anisotropic model has important consequences on the relation between the
Kondo temperature and the ``bare'' model parameters (effective bandwidth,
exchange constants $J_\perp$ and $J_z$), and on the coupling of the
effective impurity spin with the external magnetic field.

\appendix

\section{Analysis of the non-Fermi-liquid fixed point}
\label{secanalysis}

We study the properties of the non-Fermi-liquid fixed point in the $S=1$
Kondo model with $D<0$ in the extreme anisotropic limit, $|D| \to \infty$
(numerical results, where needed, are taken from a $D=-100W$ calculation).

We show that in the $N \to \infty$ limit ($N$ being the NRG iteration
number), the fixed point corresponds to fermions with residual anisotropic
exchange scattering from a $S_z=\pm 1$ magnetic doublet. In other words,
there is no decoupling of the residual impurity spin, as in the isotropic
$S=1$ Kondo model \cite{koller2005}. There is furthermore no effective
potential scattering in the $|D| \to \infty$ limit, since the conduction
band electrons cannot flip the impurity spin due to the high energy barrier,
thus there can be no quasiparticle phase shift due to Kondo screening. We
may thus take the Hamiltonian for the uncoupled conduction band chain
\cite{koller2005}
\begin{equation}
H_\mathrm{band} = \xi^{-1} \Lambda^{-(N-1)/2}
\sum_{r}
\left(
E_p^{(0)} p^\dag_{r,\sigma} p_{r,\sigma}
+
E_h^{(0)} h^\dag_{r,\sigma} h_{r,\sigma}
\right),
\end{equation}
where $p$ and $h$ are second quantization operators for particle and hole
excitations, while $E_p^{(0)}$ and $E_h^{(0)}$ are their energies in the
units of the characteristic energy scale at the $N$-th NRG iteration, and
$\xi=\ln \Lambda \sqrt{\Lambda}/(1-\Lambda^{-1}) \approx 2.41$ is an
additional scale factor in the discretization scheme used in this work
\cite{campo2005}.

The fixed-point Hamiltonian can be written as $H'=H_\mathrm{band} + H_c$
with
\begin{equation}
H_c=J^*_z s_z S_z + J^*_\perp \left( s_x S_x + s_y S_y \right),
\end{equation}
where $s_i$ are the spin operators on the first site of the Wilson chain,
\begin{equation}
s_i = \sum_{\alpha\alpha'} f^\dag_{0,\alpha} 
\left(\frac{1}{2}\sigma^i_{\alpha\alpha'}\right)
f_{0,\alpha'},
\end{equation}
while $S_i$ are the impurity spin-$1$ operators. The $f_0$ operators need to
be written in terms of the single particle and hole operators
\cite{wilson1975}:
\begin{equation}
f^\dag_{0\sigma} = \sum_r \alpha_{0r} \left(
p^\dag_{r,\sigma} + h_{r,\sigma}
\right).
\end{equation}
In the $|D|\to\infty$ limit, the Hilbert space is restricted to the $S_z=\pm
1$ states, thus the transverse exchange $J^*_\perp$ drops out. Keeping only
the terms involving the particle excitations, we finally obtain
\begin{equation}
H_c = J^*_z \sum_{r,r'} \alpha_{0r} \alpha_{0r'} S_z
\frac{1}{2} \left(
p^\dag_{r,\uparrow} p_{r',\uparrow}
- p^\dag_{r,\downarrow} p_{r',\downarrow}
\right).
\end{equation}

\begin{table}[htb]
\centering
\begin{ruledtabular}
\begin{tabular}{@{}llll@{}}
Energy & $Q$ & $S_z$ & Degeneracy \\
\colrule
0        & $0$ & $\pm 1$ & 2 \\
0.672864 & $\pm 1$ & $\pm 1/2$ & 4 \\
0.820964 & $\pm 1$ & $\pm 3/2$ & 4 \\
1.345728 & $0$ & $0$ & 2 \\
1.493829 & $\pm 2, 0 [\times 2]$ & $\pm 1$ & 8 \\
1.641929 & $0$ & $\pm 2$ & 2 \\
\colrule
\end{tabular}
\end{ruledtabular}
\caption{Excitation spectrum for large negative $D$
(here $D=-100W$, $\Lambda=2.5$, $N=79$).
The lowest single-particle excitations for free electrons is
$E_{p,1}^{(0)}=0.7468559$.
}
\label{table1}
\end{table}

The fixed-point excitation spectrum computed using NRG is shown in
Table~\ref{table1}. The ground state has total charge $Q=0$ and consists of
two spin states $S_z=\pm 1$. There are four low-lying single-particle
excited states with $Q=+1$: two degenerate levels with $S_z=\pm 1/2$ and
(scaled) energy
\begin{equation}
E_p(A) = E_{p,1}^{(0)} - \frac{1}{2} J^*_z \alpha_{01}^2 \Lambda^{(N-1)/2} \xi
\end{equation}
and two other degenerate levels with $S_z=\pm 3/2$ and energy
\begin{equation}
E_p(B) = E_{p,1}^{(0)} + \frac{1}{2} J^*_z \alpha_{01}^2 \Lambda^{(N-1)/2} \xi
\end{equation}
In the first two levels, the impurity spin and the spin of the particle
excitation are anti-aligned, while in the second two levels the impurity
spin and the spin of the particle are aligned.

Since the energy difference between $E_p(A)$ and $E_p(B)$ no longer varies
with $N$ at the fixed point and the factor ${\overline \alpha}_{01}^2=
\alpha_{01}^2 \Lambda^{(N-1)/2}$ reaches its limiting value (equal to
$0.309$ for $\Lambda=2.5$) this implies that $J^*_z$ is a constant given by
\begin{equation}
J^*_z = \frac{E_p(B)-E_p(A)}{\overline{\alpha}_{01}^2 \xi}.
\end{equation}
For $\rho J=0.1$ and $D=-100W$ (Table~\ref{table1}) we obtain
\begin{equation}
J^*_z \approx 0.198 W = 0.99 J.
\end{equation}
As expected, we obtain $J^*_z$ which is essentially equal to the bare $J$ in
the large-$|D|$ limit.

\begin{acknowledgments}
We thank A. Heinrich for sharing unpublished results for tunneling spectra
of Co impurities. We acknowledge support by Gesellschaft f\"ur
wissenschaftliche Datenverarbeitung (GWDG) in G\"ottingen through SFB 602
(T. P. and R. \v{Z}.) and project PR 298/10-1 (R.P.).
\end{acknowledgments}

\bibliography{aniso}

\end{document}